\documentclass{emulateapj}

\received{}
\revised{}
\accepted{}

\newcommand{\chandra}{{\sl Chandra}}
\newcommand{\xmm}{{\sl XMM}-Newton}
\newcommand{\einstein}{{\sl Einstein}}

\newcommand{\kms}{$\rm km\ s^{-1}$}
\newcommand{\cmt}{${\rm cm^{-2}}$}

\shorttitle{X-ray Absorption Line Intrinsic to the BL Lac}
\shortauthors{Fang et al.}

\begin{document}

\title{Detection of A Transient X-ray Absorption Line Intrinsic to the BL Lacertae Object H~2356-309}

\author{Taotao~Fang\altaffilmark{1}, David~A.~Buote\altaffilmark{1},
  Philip~J.~Humphrey\altaffilmark{1},
  Claude~R.~Canizares\altaffilmark{2}}

\altaffiltext{1}{Department of Physics \& Astronomy, 4129 Frederick Reines Hall
, University of California, Irvine, CA 92697; fangt@uci.edu}

\altaffiltext{2}{Department of Physics and Kavli Institute for
  Astrophysics and Space Research, Massachusetts Institute of
  Technology, Cambridge, MA 02139} 

\begin{abstract}

Since the launch of the \einstein~X-ray Observatory in the 1970s, a number of
broad absorption features have been reported in the X-ray spectra of
BL Lac objects. These features are often interpreted as arising from
high velocity outflows intrinsic to the BL Lac object, therefore
providing important information about the inner environment around the
central engine. However, such absorption features have not been
observed more recently with high-resolution X-ray telescopes such as \chandra~and
\xmm. In this paper, we report the detection of a
transient X-ray absorption feature intrinsic to the BL Lac object
H~2356-309 with the \chandra~X-ray Telescope. This BL Lac object was
observed during {\sl XMM} cycle 7, \chandra~cycle 8 and 10, as part of
our campaign to investigate X-ray absorption produced by the warm-hot
intergalactic medium (WHIM) residing in the foreground large scale
superstructure. During one of the 80 ksec, \chandra~cycle 10 observations, a transient absorption feature was detected at $3.3\sigma$ (or 99.9\% confidence level, accounting for the number of trials), which we
identify as the \ion{O}{8} K$\alpha$ line produced by an absorber
intrinsic to the BL Lac object. None of the other 11 observations
showed this line. We constrain the ionization parameter ($25 \lesssim \Xi \lesssim 40$) and temperature ($10^5 < T < 2.5\times10^7$ K) of the absorber. This absorber is likely produced by an
outflow with a velocity up to $1,500$ \kms.\ There is a suggestion of possible excess emission on the long-wavelength side of the absorption line; however, the derived properties of the emission material are very different from those of the absorption material, implying it is unlikely a typical P Cygni-type profile.

\end{abstract}


\section{Introduction}

\begin{figure*}[t]
\center
\includegraphics[width=1\textwidth,height=.3\textheight,angle=0]{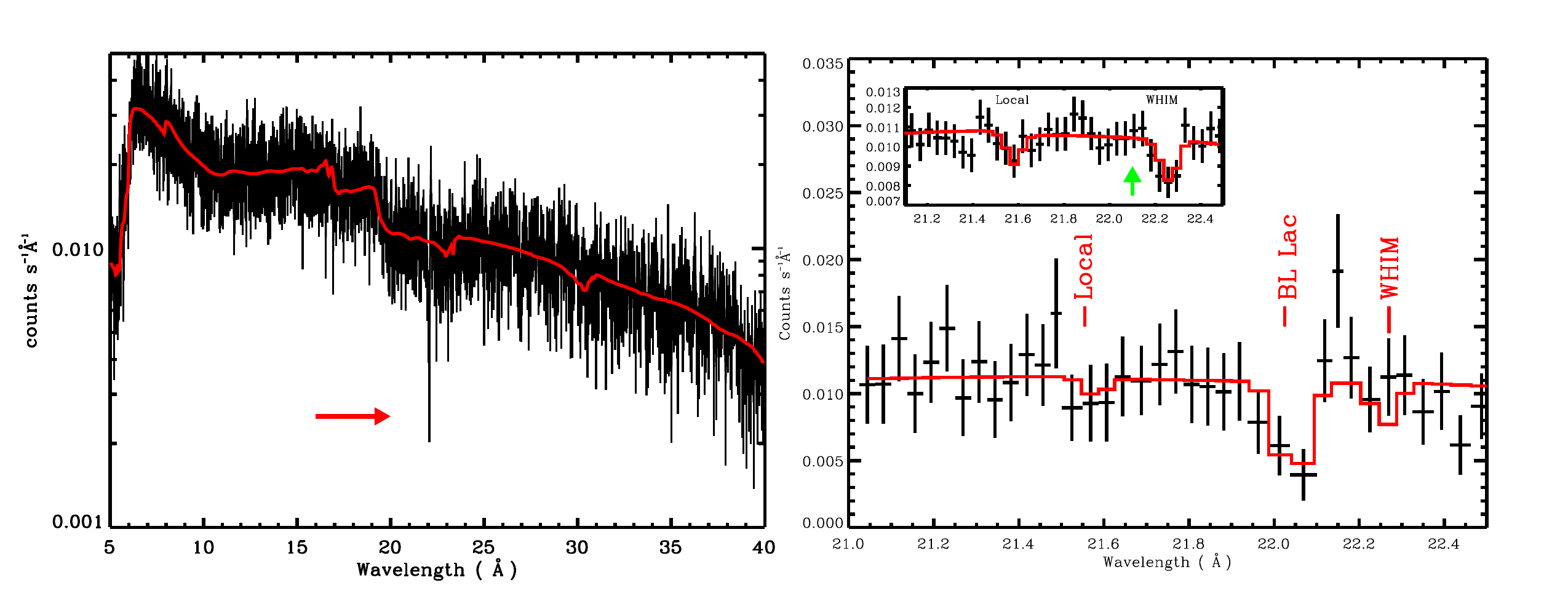}
\caption{The X-ray spectrum of H~2356-309, taken from the observation \#10498, plotted in the observer's frame. Left panel shows the entire spectrum between 1 and 40 \AA. The red line is a simple fit with a power law and the Galactic absorption. The 22.05 \AA\ feature (indicated by a red arrow) is clearly visible. Right panel shows the spectrum between 21 and 22.5 \AA.\ Red
  line is the model with three absorption features, one at 21.6 \AA\ (local Galactic 
  \ion{O}{7} $K_{\alpha}$ absorption),  one at 22.05 \AA\ (\ion{O}{8}
  $K_{\alpha}$ absorption intrinsic to the BL Lac), and one at 22.3 \AA (redshifted \ion{O}{7} $K_{\alpha}$ absorption associated with the WHIM in the Sculptor Wall). We fixed the Galactic and WHIM lines at the values measured in F10. Also shown in the
  inset is the stacked spectrum of H~2356-509 from the other ten
  Chandra observations (see F10). The two red absorption lines are the local and
  the Sculptor Wall (WHIM) \ion{O}{7} $K_{\alpha}$ lines (see
  F10). The green arrow indicates the wavelength of the transient
  feature seen in the observation \#10498.}
\label{f:abs}
\end{figure*}

Blazars, characterized by their highly polarized emission in the optical band 
and strong variability at almost all frequencies, are often
interpreted as active galactic nuclei (AGNs) with relativistic jets
beamed toward us (see, e.g., \citealp{ang80}). BL Lac objects, which
are a sub-class of blazars, typically exhibit weak or no spectral
features in emission or absorption at all wavelengths (e.g., 
\citealp{urr95}). In particular, the very few weak absorption features
detected in the optical band are believed to originate in the
interstellar medium of the host galaxy (e.g., see
\citealp{sba05,plo10}) and have been used to determine the redshift of
the BL Lac object. Therefore, unlike the typical warm absorbers seen
in AGNs, optical absorption lines in BL Lac objects offer no
information about the immediate environment of the central black
holes.

However, in the X-ray band, \citet{can84} reported the first detection
of an absorption feature in the spectrum of the BL Lac object
PKS~2155-304, using the objective grating spectrometer on the {\sl
Einstein Observatory}. Since then a number of X-ray absorption
features have been reported (see, e.g.,
\citealp{urr86,mad91,gra97,sam97}), leading to the conclusion that
such X-ray absorption features are quite common in the spectra of BL
Lac objects. These features were typically broad (with a width of a
few tens of eV up to a few hundred eV) in the soft X-ray band, and were
often interpreted as resonant absorption from highly ionized oxygen
originating in a high velocity outflow (up to a few 10,000 $\rm km\ s^{-1}$) intrinsic to the BL Lac object
(e.g., \citealp{kro85}). These discoveries demonstrate that X-ray
absorption features can provide an extremely valuable probe of the central region of BL Lac objects.

Since the launch of the \chandra~and \xmm~X-ray telescopes, a number of BL
Lac objects have been observed with unprecedented high spectral
resolution. However, so far no {\it intrinsic} X-ray absorption lines
have been detected. {\it Non-intrinsic} X-ray absorption features have been reported
in these BL Lac observations. But unlike previously detected features, when
observed with high-resolution these features are typically narrower
(width of a few eV or less) and often attributed to the
foreground Galactic (e.g., \citealp{nic02,fan03,ras03}) or
intergalactic origins (e.g., \citealp{fan02,nic05b}; \citealp{buo09}
-- B09 hereafter;
\citealp{fan10} -- F10 hereafter).
\citet{blu04} and \citet{per05} examined a number of bright BL Lac objects with {\sl XMM}-Newton. They did not detect any broad
features and argued the previous detections were affected by poor spectral quality, calibration uncertainties, as well as the simplification of the continuum model.  Although in \citet{blu04} they found a few highly significant
features (more than expected from statistic fluctuations), they
were not able to find plausible identification of them, casting
doubt on the existence of any absorption lines intrinsic to BL Lac
objects.

In this paper, we report the serendipitous detection of a transient absorption feature
during our multiple observations of the BL Lac object H~2356-309 with
gratings on board the \chandra~and \xmm~X-ray telescopes. The
primary science goal was to study the narrow absorption features
produced by the warm-hot intergalactic medium (WHIM) along the sight
line toward the BL Lac object. We clearly detected an \ion{O}{7} absorption
line produced by the WHIM in the Sculptor Wall, a superstructure along the sight
line at $z\sim0.03$ (B09, F10). During one of the exposures (observation \#10498), a strong
absorption feature was identified at $\sim 22.05$ \AA.\ None of the
other 11 \chandra~and \xmm~observations showed this feature. In this paper we discuss
several possibilities of the origin of this transient feature, and
conclude it is unlikely an instrumental feature. The most likely
explanation is an intrinsic, transient feature produced by
hydrogen-like oxygen. We also discuss the constraints on the
temperature and ionization structure. 

\section{Data Analysis}

H~2356-309 is a BL Lac object located at $z=0.165\pm0.002$
\citep{fal91}. Multi-wavelength observations of this target showed its
broad-band spectrum can be well described by the synchrotron
self-Compton emission from the relativistic jet (e.g., \citealp{hes10}). Its sight
line passes through a large-scale superstructure of galaxies, the
Sculptor Wall, at $z\sim0.03$ (see Figure 1 of B09).  With \xmm~it was observed in 2007 for approximately 130
ksec (ObsID 0504370701; see B09). With \chandra~it was observed first in 2007
during cycle
8 for 100 ksec, and then again in 2008 during cycle 10 in ten separate
exposures totaling 500 ksec. The \chandra~exposures range from $\sim$ 15 to 100
ksec (see Table 1 of F10). 

Observation \#10498 was performed on September 22nd, 2008, for 80
ksec. As in B09 and F10, we followed the standard procedures to extract the
spectra. We used the software package CIAO (Version 4.0\footnote{see
  http://asc.havard.edu/ciao}) and calibration database CALDB (Version
3.5\footnote{see http://asc.havard.edu/caldb}) developed by the {\sl
  Chandra} X-ray Center. We refer readers to B09 and F10 for details
of data extraction, and only want to emphasize a few issues
here. First, we have generated our own type II pha file, rather than
using the file produced by the standard pipeline (Reprocessing III),
to take advantage of an improved background filter not yet available
in the standard processing (Wargelin et al. 2009)\footnote{see
  http://cxc.harvard.edu/contrib/letg/GainFilter/software.html.}. Secondly,
to account for the high-order contributions of the LETG-HRC, we built
a combined response matrix to include the first to the sixth-order
contributions (see B09 and F10 for details). Finally, we rebinned the
spectrum so that we have at least 40 counts per bin to enhance the
spectral signal-to-noise ratio. We fitted the continuum with a model
that includes a power law and the Galactic neutral hydrogen
absorption and found this simple model is adequate in describing the
overall broadband spectrum. For the observation \#10498, we found a power law photon index of
$\Gamma=1.784\pm0.027$, and a 0.5 --- 2 keV flux of $1.94 \times
10^{-11}\rm\ ergs\ cm^{-2}s^{-1}$ (see F10 for details).
Unless otherwise noted, Errors are quoted at 90\%
  confidence level throughout the paper.\\

\section{Modeling}

\subsection{Intrinsic Absorption}

In Figure~\ref{f:abs} left panel we plot the X-ray spectrum of H~2356-309 between 1 and 40
\AA\ for the observation \#10498, in the observer's frame. In the right panel we show the enlarged portion between 21 and 22.5\AA.\ An absorption feature is prominently located at $\sim 22.05$ \AA.\ In the inset we show the stacked spectrum of the remaining nine \chandra~observations, and indicate the wavelength of this feature, which was not detected, with a green arrow. In this inset the absorption feature seen at $\sim 22.3$ \AA\ is an \ion{O}{7} K$\alpha$ absorption line produced by the WHIM gas in the Sculptor Wall (see B09 and F10). There is no known
instrumental feature near this feature (\chandra~Proposers'
Observatory Guide, or POG\footnote{See
http://cxc.harvard.edu/proposer/POG/}). We examined both plus and
minus orders and this feature is present in both sides with similar
strength. The total exposure time of this observation is $\sim$ 77 ksec. We also checked the consistency by splitting the exposure into two 38 ksec exposures, and we found this feature is consistently present in both exposures. We also checked the background spectrum and did not find any anomaly at this location that may have caused such an absorption feature. Considering also the transient nature of this feature, we
conclude that it is not instrumental in origin. 

With the assumption that this feature is intrinsic to H~2356-309, we
examine the possible ion species based on a combination of chemical
abundance and line strength (the oscillator strength $f$). Giving the
detected wavelength, and assuming a very generous velocity range
($\pm30,000$ \kms), the likely line transitions are \ion{O}{7}
$K_{\beta}$ at $\lambda_{rest}=18.63$ \AA,\ \ion{Ca}{18} at
$\lambda_{rest}=18.70$ \AA,\ \ion{Ar}{15} at
$\lambda_{rest}=18.82$ \AA,\ \ion{O}{8} at
$\lambda_{rest}=18.97$ \AA,\ and \ion{Ca}{17} at
$\lambda_{rest}=19.56$ \AA.\ Here we select ion species with $f>0.1$
only. Considering that both calcium and argon are orders of magnitude less
abundant than oxygen, and we did not detect the corresponding \ion{O}{7}
$K_{\alpha}$ transition, the most likely candidate is an intrinsic
\ion{O}{8} $K_{\alpha}$ absorber in an outflow.  

We fitted the spectrum of the sequence \#10498 with a model that includes the following
components: (1) Galactic neutral hydrogen absorption with a fixed
column density of $N_H = 1.33\times 10^{20}$ \cmt~\citep{dic90}; (2) a power law; (3) the WHIM and Galactic \ion{O}{7} $K_{\alpha}$
absorption lines at 21.6 and 22.3 \AA, respectively (B09 and F10);\ and (4) an intrinsic absorption line at
$\sim$ 22.05 \AA.\ We fixed the component (3) - the Galactic and the WHIM absorption lines - at the values obtained in F10. We chose the Voigt-profile
based model that was described in B09 and F10 to fit the absorption
feature (4); however, the exact form of the absorption line model is not
important here, as long as the model can provide an adequate
description of this feature. We limited the redshift range of this feature to
account for an outflow velocity within $10,000$ \kms,\ since most
observed outflows from AGNs have velocities in the range of a few
hundred to a few thousand \kms~\citep{cre03}. We performed the fit by minimizing the
$C$-statistic, which is identical to maximizing the Poisson likelihood
function \citep{cas79}, and yields less biased best-fitting parameters than the standard $\chi^2$ implementation (see \citealp{hum09} for details).

Figure~\ref{f:abs} shows the \#10498 
spectrum (black) and the fitted model (red). The 22.05 \AA\ line has an equivalent width
(EW) of $70.5\pm20.5$ m\AA.\ We are not
able to constrain the upper limit on the \ion{O}{8} $K_{\alpha}$ column density, and
obtain a best-fit value of $7.6 \times 10^{17}\rm\ cm^{-2}$ , with a
lower limit of $6.1 \times 10^{16}\rm\ cm^{-2}$. The absorption
feature is at a redshift of $z=0.163\pm0.001$ relative to the observer system. We also found a Doppler-$b$ parameter
of $278.8^{+584.8}_{-159.6}$ \kms. Thermal broadening can at most
provide a $b$ parameter of up to $100$ \kms~(for hot gas with
temperatures up to a few $10^7$ K), indicating velocity
gradient along the sight line plays a significant role in line broadening. If let free, both the Galactic line at 21.6 \AA~and the WHIM line at 22.3 \AA~are statistically insignificant due to low photon counts. This is consistent with what we found in the other Chandra observations (see Figure~2 of F10): both lines are weak in each individual spectrum and can only be detected when all the observations are analyzed simultaneously.

An accurate measurement of the BL Lac redshift is necessary to
determine the outflow velocity. Based on the measurement of the
optical absorption lines produced by the interstellar medium in the
host galaxy, \citet{fal91} determined the redshift of H~2356-309 is
$z=0.165\pm0.002$. We find the measured redshifted of the \ion{O}{8}
$K_{\alpha}$ absorption line is very consistent with that of H~2356-309, with an outflow velocity of at most $1500$ \kms.

We used Monte-Carlo simulations to assess the significance of this transient line. We fitted the \#10498 spectrum with a model that does not include the intrinsic line. When comparing with the fit that includes the intrinsic line, we found a decrease in the $C$-statistic of $\Delta C_{obs} = 21.3$. This intrinsic line was discovered when we studied the 21 to 22.5 \AA~spectral region of 12 observations (11 \chandra~and one \xmm). For the Monte-Carlo simulation in each trial we made 12 mock spectra in this spectral region to mimic the 12 observations. Specifically, each mock spectrum was made with the model obtained from the real observation (see F10 for model parameters) but without any intrinsic line, i.e., we used the same power law index, normalization, exposure time, and two absorption lines (the Galactic and the WHIM) for that observation. We then searched each mock spectrum between 21 and 22.5 \AA~to identify any negative feature that could give a decrease of $\Delta C$ equal or larger than $\Delta C_{obs}$. We ran a total of 40,000 trials, and found for 38 trials, there is at least one mock spectrum with a change in $\Delta C$ that is equal to, or greater than what was observed. This indicates a detection significance of $3.3\sigma$, or 99.9\% confidence level, accounting for the number of "trials".

When evaluating the detection significance we also fixed the Galactic and the WHIM absorption lines at the values obtained in F10. In principle, the two line parameters should be determined by a joint fit of all the 12 observations. However, such joint fit in our Monte-Carlo simulation is extremely computationally intensive, and our estimate indicated that the change in $\Delta C$ is negligible. Therefore, we decided to fix these two line parameters in our calculation.

We reiterate that the 21---22.5 \AA\ range is the appropriate wavelength
region over which to perform the random trials to assess the statistical
significance of the 22.05 \AA\ line, because it was only from examining this
limited wavelength range that, by chance, we discovered this transient
line while studying the Sculptor WHIM in F10. However, for illustrative purposes only, we also computed the
significance of this line by performing random trials over the entire
1---40 \AA\ range and  obtained a significance of 2.5$\sigma$, or 99.0\%
confidence. For comparison, it is worth noting that if we search the
entire 1---40 \AA\ range for other features in the 12 spectra, the strongest features we find
are 5 candidate lines where the decrease in the C-statistic is greater
than 10, with a maximum change of 15 ($\sim 1.7\sigma$). These candidates are even less
significant than the 22.05 \AA\ line and, importantly, none of them are
associated with ion species (with strong oscillator strength)
appropriate for the Milky Way, the blazar, or the Sculpor Wall WHIM
absorber.

\begin{figure}[t]
\centerline{\includegraphics[scale=0.35,angle=180]{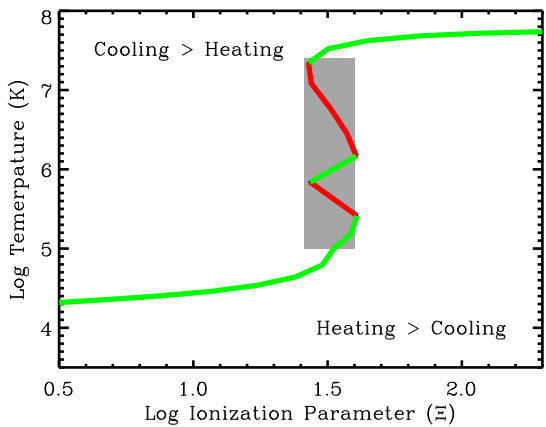}}
\caption{Thermal equilibrium curve. Green parts indicate stable
  states, while red parts indicate unstable states. The grey area
  shows the allowed region given by constraints on the ionization
  fractions.}
\label{ionpara_t}
\end{figure}

\begin{figure*}[t]
\center
\includegraphics[scale=0.32,angle=180]{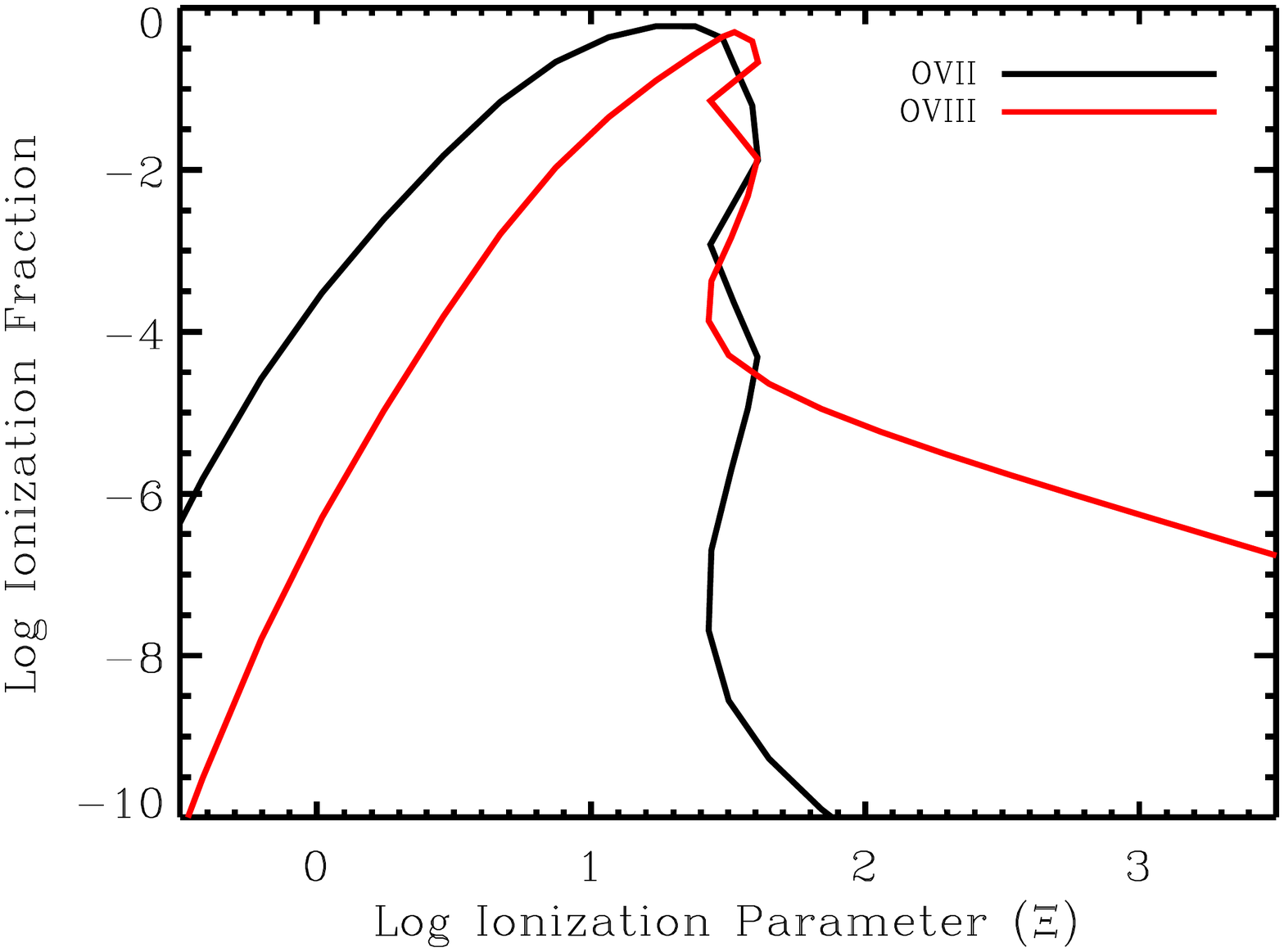}
\includegraphics[scale=0.32,angle=180]{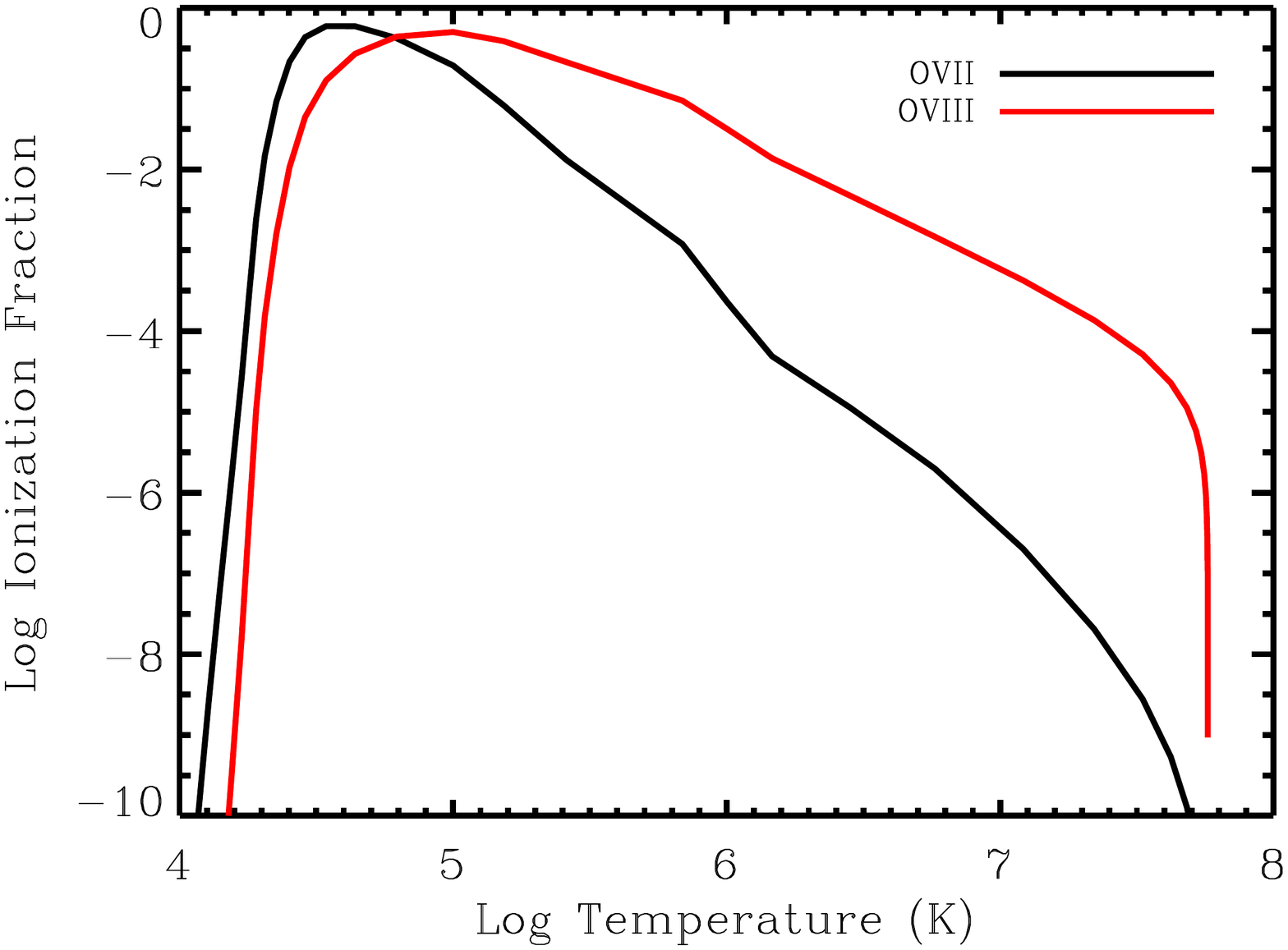}
\caption{top panel: the ionization fraction of \ion{O}{7} (black line) and \ion{O}{8} (red
line) as a function of the ionization parameter $\Xi$; bottom
panel: the ionization fraction of \ion{O}{7} (black line) and \ion{O}{8} (red
line) as a function
of temperature.}
\label{frac}
\end{figure*}

\subsection{Physical Properties}

Considering the redshift of the \ion{O}{8} $K_{\alpha}$ absorption
line, photonization by
the central black hole of the blazar H~2356-309 likely plays a major role in
ionizing the absorber. Therefore, we have used the photonization code
CLOUDY to determine its physical condition. Calculations were
performed with version 06.02 of CLOUDY, last described by
\citet{fer98}. 

In general, photo-ionized gas achieves thermal equilibrium by
balancing heating with cooling, where the major heating source is the
ionizing photons from the central black hole, and the major cooling
mechanism is collisionally excited, atomic and ionic line emission. At
high temperatures, heating by Compton scattering and cooling by
thermal bremsstrahlung radiation and inverse Compton scattering will become important. Taking all these processes into consideration, we
calculated the thermal equilibrium temperature as a function of the
ionization parameter $\Xi$, using CLOUDY (see
Figure~\ref{ionpara_t}). Following \citet{kro81}, this 
ionization parameter is defined as

\begin{equation}
\Xi \equiv \frac{L_{i}}{4\pi R^2 n_H ckT}
\end{equation}
where $L_{i}$ is the luminosity of ionizing photons, $R$ is the
distance of the absorber to the central source, $n_H$ is the gas
density, $k$ is the Boltzmann constant, and $T$ is the gas
temperature \footnote{The other commonly used definition of the ionization parameter is $\xi \equiv \left(L_i/n_HR^2\right)$. The conversion between this two definitions (in c.g.s unit) is: $\left(\xi/\Xi\right) \approx 52\ T_6$, where $T_6$ is temperature in units of $10^6$ K.}. For simplicity, we adopted a power law spectrum with a
photon index of $\Gamma = 1.784$, obtained from our \chandra~spectrum,
and also solar metallicity. We will discuss the impact of these
choices later.

In Figure~\ref{ionpara_t}, cooling dominates over heating above the thermal equilibrium curve, and heating exceeds cooling below the curve. Along the equilibrium
curve, the gas is thermally stable in the green parts, and unstable in the 
red parts where the gradient becomes negative. In the unstable
states a small increase in temperature will lead to regions where
heating exceeds cooling and therefore becomes unstable. The stable
states include one ``cold'' ($T\leq 10^5$ K), one ``hot'' ($T > 10^7$
K), and one intermediate state ($T \sim 10^6$ K). 

We also calculate the ionization fraction of both \ion{O}{7}
and \ion{O}{8}, following \citet{kro85}. The top panel of Figure~\ref{frac} shows the
ionization fraction of \ion{O}{7} (black line) and \ion{O}{8} (red
line) as a function of the ionization parameter $\Xi$; and the bottom
panel of Figure~\ref{frac} shows the ionization fraction as a function
of temperature. We do not detect the intrinsic \ion{O}{7} $K_{\alpha}$
line, and estimate a 3$\sigma$ upper limit of the line equivalent
width of $24$ m\AA.\ This puts a tight lower limit of $\Xi \gtrsim 25$,
and $T \gtrsim 10^5$ K. On the
other hand, the derived \ion{O}{8} column density is about a few
$\times 10^{17}$ \cmt.~It is therefore highly unlikely that the ionization
fraction of \ion{O}{8} is much smaller than $10^{-4}$ as this would imply
a hydrogen column density much higher than $10^{24}$ \cmt~even for
solar abundance. This puts a tight constraint on the upper limit of  $\Xi \lesssim 40$,
and $T \lesssim 2.5\times 10^7$ K. Figure~\ref{ionpara_t} shows this
allowed region in grey. 

The exact shape of the thermal equilibrium curve depends on
assumptions such as the photon index of the incident spectrum and
the metal abundance (see, e.g., \citealp{rey95}). A steeper spectrum
(e.g., $\Gamma > 3$) will lower the Compton temperature at which
Compton heating and cooling balance each other, therefore lowering the
temperature of the ``hot'' stable state in Figure~\ref{ionpara_t}. On
the other hand, a change in the metal abundance will also result in
a change in the peak positions of the thermal equilibrium curve because
of the metal line cooling mechanism. However, our estimates indicate
unless these assumptions change dramatically, they do not have
significant impact on the estimated parameters (ionization parameter,
temperature, ionization fractions, etc.) here.

\subsection{Transient Nature of the Absorber}

The observation \#10497 was taken immediately before this observation and ended on September
20th, 2008 at about 10AM; and the observation \#10762 was taken immediately after this
observation and started on September 25th, 2008 at about 2AM. This
suggests the transient feature lasts at most $t_{max} \approx 4\times 10^5$
seconds, and at least $t_{min} = 8\times 10^4$ seconds. 

Line variability is fairly common in the soft X-ray spectrum of
AGNs. In particular, recent observations of AGNs with high resolution
spectroscopy indicate narrow absorption lines can appear and vanish in
time scales less than a few 100 ksec (e.g., \citealp{gib07}). There
are two likely scenarios that an absorption line can become transient:
(1) the ionization structure of the absorber changes (see, e.g.,
\citealp{hal84}); or (2) the absorbing material changes, e.g., moving in and out of the sight
line (see, e.g., \citealp{fab94}). We consider both scenarios in the
following discussion.

To change the physical state of the absorber during such a short
period, either the ionizing source varies rapidly, or the absorber is
in a physically unstable state. The source flux is extremely stable
during our 500 ksec \chandra~observations that
span about four months (it varied
at most about 30\%; see F10). Furthermore, one \chandra~and one
\xmm~observation performed about one year before these \chandra~observations showed variations about a factor of less than 2 (B09). Hence, the
source variation is unlikely to be the cause of this transient feature. 

Considering the possibility that the absorber becomes thermally unstable (the red parts
in Figure~\ref{ionpara_t}), this intrinsic instability can lead to
the transient nature of the absorber. In this case, the ionization
structure can change rapidly if the photonization timescale is longer
than the time interval $t_{min}$. This photonization timescale can
be estimated as 

\begin{equation}
t_{ion} = \left[\int_{\nu_{th}}^{\infty} \frac{L_{\nu}\sigma(\nu)}{4\pi
    R^2 h\nu} d\nu\right]^{-1}.
\end{equation}
Here $L_{\nu}$ ($\propto \nu^{-\alpha}$ where $\alpha=\Gamma-1$ is
the spectral index) is the ionizing photon flux, $\sigma$ is the photonization
cross section and $\propto \left(\nu_{th}/\nu\right)^3$, $\nu_{th}$ is
the photonization threshold frequency, and $h$ is the Planck
constant. Adopting the numbers for \ion{O}{8} \citep{ver96}, we found
$t_{ion} \approx 2\times 10^4 R_{pc}^{2}L_{46}^{-1}\ s$. Here $R_{pc}$ is the
distance to the absorber in units of pc, and $L_{46}$ is the
ionizing luminosity in units of $10^{46}\rm\ ergs\ s^{-1}$. For H~2356-309,
$L_i$, the luminosity of the ionizing photons, is $\sim 5\times10^{45}\rm\ ergs\ s^{-1}$. If $t_{ion} \gtrsim t_{min}$, we
found the distance of the absorber must be $R \gtrsim 3$ pc. This
distance would put the absorbing material somewhere between the
typical broad line region (BLR, sub-pc) and narrow line region (NLR,
10 pc -- 1 kpc) of an AGN. The density of the absorber then is

\begin{equation}
n_H \approx 7 \times 10^5 R_{pc}^{-2} T_{6}^{-1} \Xi_{30}^{-1}
L_{46}\ \rm cm^{-3},
\end{equation}
where $T_6$ is the temperature in units of $10^6$ K, and $\Xi_{30}$ is
the ionization parameter in units of 30. Taking the typical values for
H~2356-309, the density is $n_H \approx 4\times 10^5\rm\ cm^{-3}$. With this
density, the typical recombination time scale, $t_{rec} \approx 4
\times 10^6\ T_6^{1/2}(n_H/10^5\rm\ cm^{-3})^{-1} \approx 10^7$ seconds
, is also longer than $t_{min}$.  

If instead the absorber is stable but moves in and out of the sight
line between observations, then all the time scales must be shorter
than $t_{max}$. The absorber then has an upper limit on the distance
of $R\lesssim 6$ pc (from $t_{ion} < t_{max}$), and a lower limit on
the density of $n_H \gtrsim 10^6\rm\ cm^{-3}$ (from $t_{rec} <
t_{max}$). The \ion{O}{8} column density is $\sim 10^{17}\rm\
cm^2$. Assuming solar metallicity and an ionization fraction of 0.5,
the size of the absorber along the sight line is $\sim 3 \times
10^{13}\rm\ cm$. If the absorber has a similar size in the
perpendicular direction, it implies a light crossing time of $\sim
3\times10^5$ seconds if the velocity in that direction is $\sim
1,000\rm\ km\ s^{-1}$. Since this time is less than $t_{max}$, this scenario is also consistent with the data.

We can also estimate the mass loss rate assuming a uniform spherically
symmetric outflow \citep{kro85}:

\begin{eqnarray}
\dot{M} & = & f\frac{m_H \upsilon L_{i}}{ckT\Xi} \nonumber \\
        & \approx & 15 f_{0.1} \ L_{45}T_6^{-1}\upsilon_{1000}\Xi_{30}^{-1}\rm\ M_{\odot}yr^{-1},
\end{eqnarray} 
Here
$m_H$ is the hydrogen mass, $\upsilon_{1000}$ is the outflow velocity
in units of 1,000 \kms, and $f$ is the covering fraction and is
roughly $0.1$ --- the fraction of observational time showing the
line. This value is higher than those of Seyfert galaxies which
typically have weak outflows ($\sim 1 \rm M_{\odot}yr^{-1}$, e.g.,
\citealp{ste05}), but comparable to those AGNs with energetic jets
(e.g., \citealp{ste09}).  The mass loss rate is likely smaller if it
is beamed. Such a radial outflow can produce a P Cygni-type
line profile typically seen in a stellar wind
\citep{kro85}. Interestingly, we do obtain a marginal detection of an emission feature on the long-wavelength side of the absorption feature, as expected from
the P Cygni profile (see discussion below.)

\section{Discussion}

\begin{figure}[t]
\center
\centerline{\includegraphics[scale=0.33,angle=180]{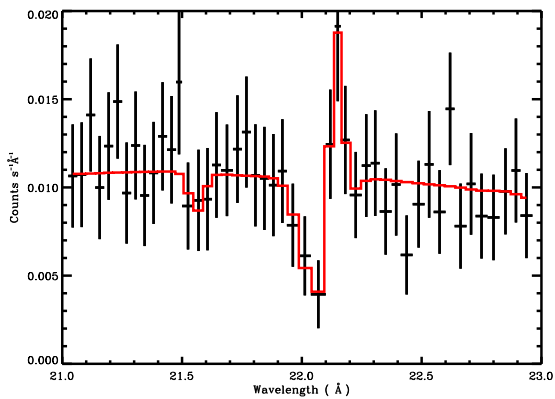}}
\caption{Spectral fitting with a P Cygni-type profile, which includes
  one absorption line model on the short-wavelength side, and one emission line on
the long-wavelength side. The wavelength is plotted in the obseverber's frame.}
\label{pc}
\end{figure}

\subsection{A Possible P Cygni Profile?}

The P Cygni-type of line profile was originally discovered in the
optical and ultraviolet spectra of stellar objects, and is often
attributed to the stellar wind (e.g., \citealp{lam87}). With {\sl
Chandra}, the X-ray P Cygni line was first detected in the spectrum of
Circinus X-1, a Galactic X-ray binary (\citealp{bra00,sch02}) and,
subsequently, it was also found in the X-ray spectra of active
galactic nuclei (e.g,
\citealp{kas01}). A P Cygni-type of line profile provides an important
diagnostic tool of whether the outflow is beamed with a jet-like
structure or is more spherically extended.  

In the observation \#10498, we find a possible emission feature right next to the
absorption feature on the long-wavelength side, resembling a P
Cygni-type profile. We fit this emission feature with a Gaussian
profile, along with an absorption profile on the short-wavelength side (see
Figure~\ref{pc}, plotted in the observer's frame). We find this is sufficient for fitting the
absorption-emission feature at $22.05$ \AA.\ Using Monte-Carlo
simulations as described in \S~3.1, we found this feature is detected at the $2.3\sigma$
level. If this transient emission feature is \ion{O}{8} $K\alpha$ and can be associated with the absorber, we measured an \ion{O}{8} line flux
of $8.2 \times 10^{-5}\rm\ photons\ cm^{-2}s^{-1}$. The \ion{O}{8}
line emissivity peaks at about $3\times10^6$ K. Assuming a peak
emissivity, we obtained an upper limit of the emission measure of $EM = \int n_e^2 dV
\approx 5\times 10^{10}\ \rm cm^{-6} pc^3 $ at the distance of the BL
Lac. Here $n_e$ is the electron density and the integration is over
the emission volume. However, we consider the P Cygni scenario unlikely since if the $n_e \sim 10^6\rm\ cm^{-3}$ as we estimated for the absorber, the linear size of the emission material ($\sim$ 0.4 pc) would be far greater than that of the absorber. Clearly, more sophisticated modeling is
necessary to fully understand the structure and physical properties of
this material as revealed by the emission/absorption profile.

\subsection{Host galaxy, Intervening or Local Absorption?}

The transient nature of this absorption feature makes it very unlikely to be
produced by the ISM in the host galaxy, an intervening absorber, or a local absorber. We do notice that the
observed wavelength of this feather is very close to the rest
wavelength of the \ion{O}{6} $K_{\alpha}$ inner shell transition
($\lambda = 22.02$ \AA, see \citealp{pra03,sch04}). This \ion{O}{6}
$K_{\alpha}$ inner shell transition was first reported in
\citet{lee01} in the X-ray spectrum of MCG-6-30-15.

\subsection{Summary}

X-ray observations of narrow absorption features offer a unique
opportunity to probe the inner region of BL Lac objects. In this paper
we report the detection of a transient absorption line during
our H~2356-309 campaign with the \chandra~X-ray Telescope. This line
is most likely produced by \ion{O}{8} in a photo-ionized outflow
intrinsic to the BL Lac object H~2356-309. Considering the transient
nature of the absorber, we obtain constrains on the absorber's
ionization parameter, $25 \lesssim \Xi \lesssim 40$, temperature,
$10^5 < T < 2.5\times10^7$ K, and density, a few $\times 10^5\ \rm
cm^{-3}$.

Our detection is quite different from X-ray absorption features detected in BL Lac objects before \chandra~and \xmm~(e.g., \citealp{can84,mad91}). Those absorption features typically have a velocity
width of up to a few $\times 10^4$ \kms,\ while in H~2356-309, the
velocity is at most $1-2 \times 10^3$ \kms.\ However, even in our
case, the line width is much larger than that expected from thermal
broadening, suggesting an outflow as a likely cause of the broadening.

\citet{blu04} studied the  \xmm~RGS spectra of four BL Lac objects with
previously known, broad X-ray absorption lines and found none. \citet{per05} also analyzed the X-ray spectra of 13 bright BL Lac objects observed with {\sl XMM}-Newton. They did not detect any broad, intrinsic features either, but they found strong evidence for the intrinsic curvature of the spectral index of most of the targets. In both studies, they concluded that the previously reported features were due to a combination of calibration uncertainties and the use of a overly simplified, single power-law model. At low resolution and low S/N a spectral curvature can mimic a broad  absorption if the spectrum was fitted with a single power law \citep{per05}. However, our
observation, which has none of these problems, along with their
detections of several unexplainable absorption features detected in \citet{blu04}, raise again
the question of whether or not such absorption is common in
BL Lac objects. The high variability of the BL Lac object and its
environment make it a challenge to address this issue. We detect this line
in one ($\sim 80$ ksec) observation for a total of 12 ($\sim 600$ ksec for \chandra,~and $\sim$ 130 ksec for \xmm) observations. Taking this probability at face value, a
long-term, monitoring program which focuses on several bright BL Lac objects
would be a feasible approach to unveil the nature of these transient absorption lines. \\

{\it Acknowledgments:} We thank Brad Wargelin for assistance with
observation set-up, Peter Ratzlaff for helping implement the new
filtering procedure, and Vinay Kashyap for assistance with the {\sl
Chandra} observation \#10498. We also thank Aaron Barth and H\'el\`ene Flohic for helpful discussions. T.F., D.A.B., and
P.J.H. gratefully acknowledge partial support from NASA through
Chandra Award Numbers GO7-8140X and G09-0154X issued by the Chandra
X-Ray Observatory Center, which is operated by the Smithsonian
Astrophysical Observatory for and on behalf of NASA under contract
NAS8-03060. We also are grateful for partial support from NASA-XMM
grant NNX07AT24G. C.R.C. acknowledges NASA through Smithsonian
Astrophysical Observatory contract SV1-61010.

\end{document}